\documentclass[a4paper, 12pt]{article}
\usepackage{comment}
\usepackage{amsmath, amssymb, amsthm}
\usepackage{graphicx,color}
\usepackage{graphics}
\usepackage{rotating}

\usepackage[utf8]{inputenc}
\usepackage[T1]{fontenc} 
\usepackage{lmodern} 
\usepackage{natbib} 
\usepackage{caption}

\usepackage[english]{babel}
\usepackage{booktabs}
\usepackage{amsmath,amsthm}
\usepackage{todonotes}

\usepackage{setspace}   
\usepackage[pdfborder={0 0 0}]{hyperref} 
\usepackage{threeparttable} 
\usepackage{tablefootnote}
\usepackage{enumitem}

\usepackage{booktabs}
\usepackage{pgfplots}
\usepgfplotslibrary{fillbetween}
\usetikzlibrary{patterns}
\pgfplotsset{width=10cm, compat=1.12}

\usepackage{dcolumn}    
\usepackage{adjustbox}  
\usepackage{placeins}   
\usepackage{float}      
\usepackage{authblk}    

\usepackage[margin=1in]{geometry}
\usepackage{lineno} 

\usepackage[toc,page]{appendix} 
\usepackage{subcaption} 
\usepackage{indentfirst}

\hypersetup{
colorlinks,
citecolor=blue,
linkcolor=blue,
urlcolor=blue} 
\graphicspath{{figures/}}
\author[1]{Alejandro Corvalan\vspace{0cm}}
\affil[1]{Universidad de Chile\vspace{0cm}}

\date{October 2025}

\title{The Distributional Consequences of \\
Paid-Priority Queues}

\vspace{1.6cm}

\begin{document}

\onehalfspacing

\maketitle
\begin{abstract}

This note examines the distributional implications of introducing a fast-track queue for accessing a service when agents are heterogeneous in both income and service valuation. Relative to a single free queue, I show that willingness to adopt the priority system is determined solely by income, regardless of service valuation. High-income individuals benefit from the fast-track access, while low-income individuals are worse off and remain in the free line. Middle-income individuals weakly prefer the single free queue; yet, under the priority regime, they pay for fast-track access. Thus, the use of the priority queue does not reveal preferences for the priority system. 

\end{abstract}

\pagebreak

\section{Introduction}

Non-market mechanisms for allocating goods and services are pervasive across economies. Although such systems often involve inefficiencies, they are frequently defended on egalitarian grounds. Accordingly, economists have long examined the distributional consequences of different allocation methods. Several studies analyze how individuals at different income levels are affected by markets, queues, rationing, and hybrid mechanisms that combine market and non-market elements \citep{sah1987queues,polterovich1993rationing,alexeev2001income,clark2007paying}. Overall, markets tend to favor the wealthy, while the poor benefit in more diverse ways from various forms of non-market allocations.

This note examines the distributional implications of introducing a fast-track queue for accessing a service when agents are heterogeneous in both income and service valuation. In the single-queue system, individuals wait in a line for a free pass to the service; in the priority regime, a second paid, fast-track queue is implemented. I compare who wins and loses when the priority line is introduced and assess the role of service valuation in relation to income. 

I show that willingness to adopt the priority system is determined solely by income, regardless of service valuation. High-income individuals benefit from the fast-track access, while low-income individuals are worse off and remain in the free line. Middle-income individuals weakly prefer the single free queue; yet, under the priority regime, they pay for fast-track access. Thus, the use of the priority queue does not reveal preferences for the priority system.

This work contributes to the literature discussing the welfare implications of non-market allocations \citep{weitzman1977price, condorelli2013market, dworczak2021redistribution, akbarpour2024redistributive}. The priority system can be framed within the language of mechanism design. By adjusting the price and time costs of queues, the central planner can balance the efficiency–equity trade-off in line with specific welfare objectives. In addition, the note is related to the study of queues as costly mechanisms to reveal valuations.  \citep{barzel1974theory, nichols1971discrimination, platt2009queue, chakravarty2013optimal}.

\section{The Free-Single Queue System}

The economy consists of a continuum of individuals who are heterogeneous in two dimensions: income $y \in [0,1]$ and value for a service $\theta \in [0,1]$. The distribution of incomes and values is  $F(y,\theta):[0,1]\times [0,1] \rightarrow [0,1]$. Each individual has one unit of time. 

The original utility of the individual, i.e., not using the service, is:

\begin{equation*}
U_0=v(y)+t
\end{equation*}

The service may be a theater performance or an appointment in a public hospital; the value, then, may be the preference for the show or the need for medical care. The service is indivisible, non-resalable, and individuals derive utility only from the first unit. A non-profit company offers $\rho<1$ units of the service.  

To allocate the service, the company implements a queue system. To attend, individuals must wait a time $c$ in the line, and $c$ is such that it clears the market. The utility of an individual who waits in line and uses the service is:

\begin{equation*}
U_Q=v(y)+\theta+t-c 
\end{equation*}

Consequently, an individual pays the queue waiting cost if $\theta\geq c$. The market-clearing condition is: 

\begin{equation*}
\rho=\int_0^1 \int_c^1 dF(y,\theta)
\end{equation*}

At equilibrium, all individuals who value the service more than a certain threshold wait in line and use it. The result follows from the fact that time is evenly distributed across agents. Whether the rich or the poor use the service depends on the correlation between income and values, given by $F(y,\theta)$.

\section{The Paid-Priority Queue System}

The company decides to incorporate the option to pay for shortening the queue time. The individuals have two options. First, they wait for a time $c_1$ in a queue to gain free access to the service. Secondly, to wait for a shorter time, $c_2 < c_1$, for a price $p$. This second alternative is the paid-priority or fast-track queue. The time and income prices $(c_1, c_2, p)$ are those that clear the market. The respective utility of those who choose to wait in either line and access the service is:

\begin{eqnarray*}
U_{Q_1}&=&v(y)+t-c_1+\theta \\
U_{Q_2}&=&v(y-p)+t-c_2+\theta
\end{eqnarray*}

The utility $U_{Q_1}$ is similar to $U_{Q}$ but $c_1 \neq c$ because the market-clearing condition also considers those in the priority queue. The individuals prefer $U_{Q_1}$ to $U_0$ if $\theta\geq c_1$.

To compare $U_{Q_2}$ with $U_0$, it is useful to define $\theta^*(y,p)=v(y)-v(y-p)$. Individuals who prefer to pay the shorter fast-track queue are those with $\theta^*(y,p)+c_2<\theta$. The function $\theta^*(y,p)$ is decreasing in $y$ given the concavity of $v$. Accordingly, the indifference curve between $U_{Q_2}$ and $U_0$ is decreasing in $\theta$. At equilibrium, those who access the service have both a high income and a high valuation of the service. 

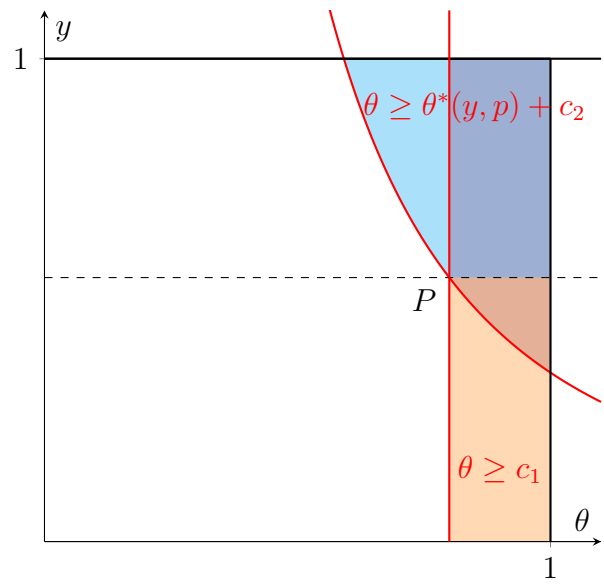
\begin{figure}[h!]
\begin{center}
\begin{tikzpicture} [xscale=0.87, yscale=1.0]
\begin{axis}[
    axis lines=middle,
    xlabel={$\theta$},
    xtick={0,1},
    ylabel={$y$},
    ytick={0,1},
    xmin=0, xmax=1.1,
    ymin=0, ymax=1.1,
    samples=200,
    domain=0.1:1.1
]

\addplot[name path=curve, red, thick, domain=0.1:1.1] {0.35/(x^2)};

\addplot[name path=yone, black, thick, domain=0:1.1] {1};
\addplot[name path=zero, domain=0:1.1] {0};
\addplot[name path=equilibrium, dashed, domain=0:1.1] {0.5468};

\addplot[cyan!30] fill between[
    of=curve and yone,
    soft clip={domain=0.5915:0.8}
];

\addplot[cyan!30] fill between[
    of=equilibrium and yone,
    soft clip={domain=0.8:1.0}
];

\addplot[orange!50!blue, opacity=0.3] fill between[
    of=curve and yone,
    soft clip={domain=0.8:1}
];

\addplot[orange, opacity=0.3] fill between[
    of=zero and equilibrium,
    soft clip={domain=0.8:1}
];

\addplot[black, thick] coordinates {(1,0) (1,1)};

\addplot[black, thick] coordinates {(0,1) (1,1)};

\addplot[red, thick] coordinates {(0.8,0) (0.8,1.1)};

\node[black] at (axis cs:0.75,0.50) {$P$};
\node[red] at (axis cs:0.90,0.15) {$\theta \geq c_1$};
\node[red] at (axis cs:0.85,0.90) {$\theta\geq \theta^*(y,p)+c_2$};

\end{axis}
\end{tikzpicture}
\caption{Allocation in the priority system}\label{fig1}
\end{center}
\end{figure}

Figure \ref{fig1} illustrates the group of individuals accessing the service for each allocation scheme.   

\begin{center}
[FIGURE \ref{fig1} ABOUT HERE]
\end{center}

In the figure, the zone in light orange includes all individuals who decide unambiguously to use the service by waiting in the free line, while those in light blue opt for the fast-track line. There are two darker zones, the intersection of the two groups, where individuals prefer either of the two alternatives for accessing the service. 

The indifferent curve $U_{Q_1}=U_{Q_2}$ is given by $\theta^*(y,p)=c_1-c_2$. It is direct to notice that point $P$, where $\theta^*(y,p)=\theta-c_2$ intersects $\theta=c_1$, is in the indifference curve. For reasons that will become clear below, I call $\underline{y}$ the income of the individual in $P$, defined implicitly by $\theta^*(\underline{y},p)=c_1-c_2$. To the right of the point $P$, the indifference curve is horizontal, since in either case individuals access the service, and so changes in $\theta$ do not change their preferences for the priority line. As a consequence, all individuals with an income above $\underline{y}$ prefer the paid priority queue (in dark blue), and all those with an income below prefer the free queue (in dark orange).  

The market-clearing condition is: 

\begin{eqnarray*}
\rho&=&\int_{\underline{y}}^1 \int_{\theta^*(y,p)+c_2}^{1} dF(y,\theta)+\int_0^{\underline{y}} \int_{c_1}^1 dF(y,\theta)
\end{eqnarray*}

The first term is the fraction of individuals in the priority queue, while the second represents those who prefer the free line. There are several price values $(c_1,c_2,p)$ that satisfy the market-clearing condition. In particular, $c_1=1$, $c_2=0$ is the pure price system. In contrast, $ c_1 = c_2 = c$ represents a single-queue system. Between these two extremes, there are multiple combinations of market and non-market systems.  

The priority system can be framed within the language of mechanism design. A central planner implements the following allocation rule: individuals located in the orange region of Figure \ref{fig1} receive one unit of the service in exchange for a time transfer $c_1$; those in the blue region receive one unit conditional on a combined transfer consisting of time $c_2$ and income $p$; individuals outside these regions receive no service. Because these allocations align with individual incentives, the mechanism is incentive-compatible over the entire domain $[0,1]\times [0,1]$, while ensuring market clearing. Importantly, by adjusting the vector of transfers $(c_1, c_2, p)$, the planner can balance the efficiency–equity trade-off in line with specific welfare objectives \citep{condorelli2013market, akbarpour2024redistributive}.

\section{Distributional Effects}

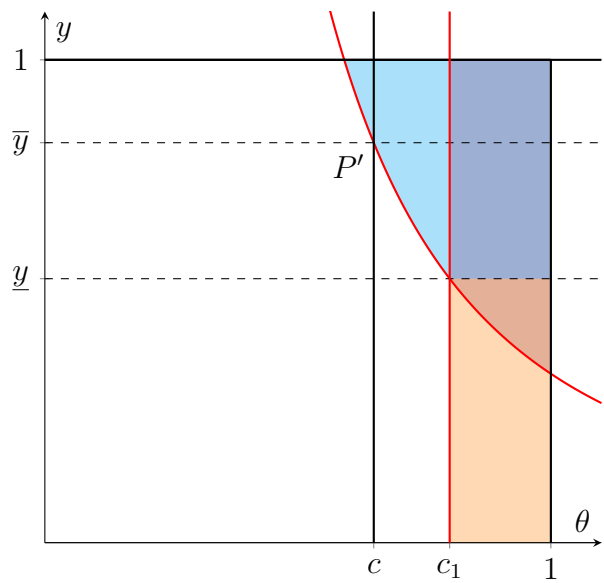
\begin{figure}[h!]
\begin{center}
\begin{tikzpicture} [xscale=0.87, yscale=1.0]
\begin{axis}[
    axis lines=middle,
    xlabel={$\theta$},
    xtick={0,0.65,0.8,1},
    xticklabels={0,$c$,$c_1$,$1$},
    ylabel={$y$},
    ytick={0,0.5468,0.8284,1},
    yticklabels={0,$\underline{y}$,$\overline{y}$,$1$},
    xmin=0, xmax=1.1,
    ymin=0, ymax=1.1,
    samples=200,
    domain=0.1:1.1
]

\addplot[name path=curve, red, thick, domain=0.1:1.1] {0.35/(x^2)};

\addplot[name path=yone, black, thick, domain=0:1.1] {1};
\addplot[name path=zero, domain=0:1.1] {0};
\addplot[name path=equilibrium, dashed, domain=0:1.1] {0.5468};
\addplot[name path=equilibrium2, dashed, domain=0:1.1] {0.8284};

\addplot[cyan!30] fill between[
    of=curve and yone,
    soft clip={domain=0.5915:0.8}
];

\addplot[cyan!30] fill between[
    of=equilibrium and yone,
    soft clip={domain=0.8:1.0}
];

\addplot[orange!50!blue, opacity=0.3] fill between[
    of=curve and yone,
    soft clip={domain=0.8:1}
];

\addplot[orange, opacity=0.3] fill between[
    of=zero and equilibrium,
    soft clip={domain=0.8:1}
];

\addplot[black, thick] coordinates {(1,0) (1,1)};

\addplot[black, thick] coordinates {(0,1) (1,1)};

\addplot[black, thick] coordinates {(0.65,0) (0.65,1.1)};

\addplot[red, thick] coordinates {(0.8,0) (0.8,1.1)};

\node[black] at (axis cs:0.60,0.78) {$P'$};



\end{axis}
\end{tikzpicture}
\caption{Allocation in the single queue and the priority system}\label{fig2}
\label{fig:fig2}
\end{center}
\end{figure}

The question addressed in this note concerns who wins and who loses when a priority system replaces a single free queue. Figure \ref{fig2} displays the set of individuals accessing the service for the two allocation schemes.  

\begin{center}
[FIGURE \ref{fig2} ABOUT HERE]
\end{center}

The shaded areas are the same as in Figure \ref{fig1}, showing all service users under priority. The black line represents the single queue system, which is simply a value threshold, with the cost of waiting satisfying the market-clearing condition. 

It is immediately noticeable that those who prefer the free queue in a priority system (in dark and light orange) are worse off than under the single queue. The reason is that the waiting time increased, given that $c_1>c$, which is a consequence of the respective market-clearing conditions. The priority system reduces the number of people in the free queue by making it more expensive. 

In addition, individuals below the curve $\theta=\theta^*(y,p)+c_2$ and between $c$ and $c_1$ can only access the service under the free-queue system; accordingly, they are strictly worse off under the fast-track alternative. Additionally, all those with $y<\underline{y}$ and $\theta<c$ are indifferent because they are not interested in the service in either case. Thus, individuals with incomes below $\underline{y}$ are weakly worse off under the priority system. 

For those individuals who would pay for the priority queue, the argument is similar to the one already made. The individual at $P'$ is indifferent between the paid-priority and the free-single queue. I call $\overline{y}$ the income of the individual in $P'$, defined implicitly by $\theta^*(\overline{y},p)=c-c_2$. As $c>c_1$, it is direct to notice that $\overline{y}>\underline{y}$ given the strict concavity of $v$. As before, changes in $\theta$ do not change the relationship, so at the right of $P'$ the indifference curve between the two allocation systems is horizontal. Accordingly, all those individuals richer than $\overline{y}$ would prefer priority, while those below that income threshold are worse off if a fast-track line is added to the single queue system. 

The following proposition summarizes these results:  

\bigskip

\textbf{PROPOSITION 1.} 

Consider a population of individuals heterogeneous in their incomes and valuations of a service. The distributional effects from allocating attendants through a free-single compared to a paid-priority queue are such that: 

\begin{itemize}
\item High-income individuals, with $\overline{y}<y\leq 1$, weakly gain with the priority system, independently of their valuations. If accessing the service, they use the fast-track line. 
\item Middle-income individuals, with $\underline{y}<y\leq \overline{y}$, are weakly worse off relative to the standard queue. However, they weakly prefer to use the priority fast-track line once it exists.
\item Low-income individuals, with $0\leq y \leq \underline{y}$, are weakly worse off under the priority system. They never use the fast-track option.
\end{itemize}

The proposition establishes that the set of individuals who are weakly better or worse off under one of the systems is solely defined by income, regardless of the valuation of the service. This sharp divide does not apply to strict preferences, as individuals with a low valuation of the service are not willing to access it across all income groups. The difference in service value does not play a role because individuals with high valuation will access the service under both allocation systems; thus, they only compare the time and income costs, disregarding the benefits they would obtain under any allocation. The same occurs with individuals who have a low valuation of the service, who will not use it. Admittedly, for inter-media valorations of service, some agents will use the service only for one allocation and not the other. However, as the proposition stated, those individuals are sharply divided by income.  

Notoriously, the priority system increases the time price of the free line to offer the service in the fast-track lane. As a consequence, the middle-income group weakly prefers the single free queue; yet, under the priority regime, they pay for fast-track access. In that sense, the use of the priority queue does not reveal preferences for the priority system.

\clearpage

\bibliographystyle{chicago}
\bibliography{bib_queues}{}

\newpage

\section*{Figures}

\end{document}